\documentclass{article}
\usepackage{spconf,amsmath,graphicx}
\usepackage[caption=false]{subfig}
\usepackage{balance}

\title{Sparse InSAR Data 3D Inpainting for Ground Deformation Detection Along the Rail Corridor}
\name{Odysseas Pappas$^1$, Juliet Biggs$^2$, David Bull$^1$, Alin Achim$^1$, Nantheera Anantrasirichai$^1$  \thanks{This work was supported by the EPSRC Impact Acceleration Account (IAA) Award: Digital Environment Sandpit. The authors would like to thank SatSense Ltd. and Network Rail Ltd. for their kind provision of data.}}
\address{$^1$Visual Information Laboratory, University of Bristol, Bristol BS8 1UB, UK \\
$^2$School of Earth Sciences, University of Bristol, Bristol BS8 1RJ, UK}
\begin{document}
%
\maketitle
\begin{abstract}
Monitoring of ground movement close to the rail corridor, such as that associated with landslips caused by ground subsidence and/or uplift, is of great interest for the detection and prevention of possible railway faults. Interferometric synthetic-aperture radar (InSAR) data can be used to measure ground deformation, but its use poses distinct challenges, as the data is highly sparse and can be particularly noisy. Here we present a scheme for processing and interpolating noisy, sparse InSAR data into a dense spatio-temporal stack, helping suppress noise and opening up the possibility for treatment with deep learning and other image processing methods. 
\end{abstract}
\section{Introduction}
\label{sec:intro}

The rail network is one of the key elements of transport infrastructure in most countries, facilitating the movement of people and goods throughout regions. Rail service unavailability can not only critically disrupt transport flows but also incur significant economic costs and even cause societal problems \cite{WOODBURN}. While disruptions in rail service can occur due to a vast number of factors, environmental or otherwise, one particularly severe can be the blocking of railway lines by soil, rocks and earth. This often occurs when a landslip or other ground movement causes a translational, rotational or other failure on an slope (such as an embankment or cutting) along the rail corridor. 

Monitoring ground movement along the rail corridor, detecting anomalous movements for a given region, and possibly forecasting ground movement that may result in rail service disruption is therefore of direct interest to organisations tasked with the operation and maintenance of rail networks \cite{CHANG}. \textit{In situ} methods (e.g. GPS, manual inspections) for monitoring ground deformation and rail network structural health are often labor and time intensive, and can only feasibly be deployed in locations where \textit{a priori} knowledge of risk of fault or deformation exists. 

Synthetic Aperture Radar Interferometry (InSAR) can provide ground deformation data at millimetre accuracy over wide regions of interest and with a revision frequency of only a few days. This makes it a prime candidate for detecting landslides \cite{Tofani}, thus complementing and enhancing railway structural health monitoring \cite{NORTH}. Such InSAR ground deformation data have in the past been used for  monitoring the structural health of rail infrastructure \cite{CHANG} \cite{NORTH}, detecting (and even predicting) uplift and subsidence in the built environment \cite{PUI1} \cite{HILL} as well as possible volcanic activity \cite{PUI2}. 

Previous work has looked at detecting and predicting ground movement, working with the complete time series data of a small number of hand-selected points of interest \cite{HILL}, and also at detecting ground movement over larger areas by converting the sparse point representation to a dense deformation image suitable for deep learning processing \cite{PUI1}. In the later case however only the average velocity, rather than the full time series data, was utilised.

Here we present a method for producing a temporal stack of dense ground movement images by conditioning, denoising and inpainting the data, in the hope of leveraging the above two approaches \cite{HILL} \cite{PUI1}. This would open up the possibility of employing various image processing methods, including deep learning architectures such as those employed in \cite{PUI1}. We present a simple proof of concept using a variance-based detector and demonstrate superior results compared to working with the original sparse time series data. It should be noted that existing deep learning-based techniques of ground movement detection are applied to wrapped InSAR data, where appearing fringes are the key features of object detection \cite{InSAR2018}. However, this requires a large area of ground movement, whist a landslip at railway embankment may be represented by only one InSAR data point.
 
\section{DATASET}
The Sentinel-1 satellite constellation provides SAR imagery of high resolution at near global coverage, with a revisit time of 6 days. This imagery can be used to form interferometric image pairs and ground motion data maps. 

There are various techniques for producing InSAR ground motion data - most fall in one of two categories, small baseline methods and persistent scatterer methods. Small baseline is suitable for large area monitoring, but is susceptible to including many incoherent points; persistent scatterer methods typically produce more reliable measurements but result in a sparse representation of the area of interest \cite{PUI1}.

Here we employ data provided by SatSense Ltd., generated using their proprietary RapidSAR pixel selection algorithm \cite{RapidSAR}. This results in interferograms that largely avoid incoherent points and produce ground motion measurements of enhanced accuracy and high SNR. These are sparse ground motion products, relying on identifying strong solid reflectors in the scene, and therefore contain unevenly spatially distributed measurement points.The built environment can provide multiple suitable reflectors, leading to high measurement density. Rural areas on the other hand present fewer such opportunities, meaning the already sparse data become even sparser, containing much fewer reliable data points in rural areas of high vegetation. 



\begin{figure}[!t]
\centering
\subfloat[]{\includegraphics[width=0.23\textwidth]{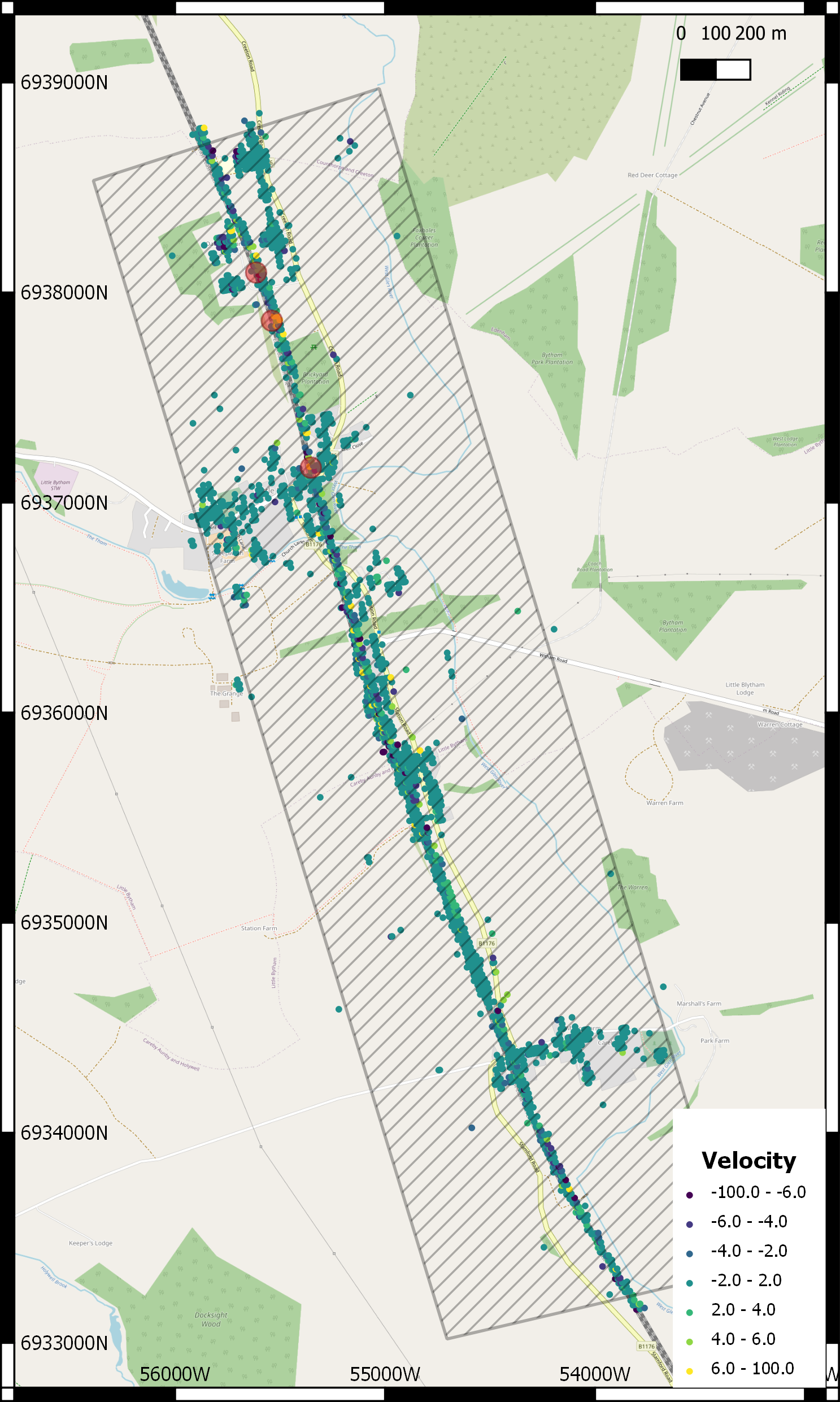}}
\hspace{0.05cm}
\subfloat[]{\includegraphics[width=0.23\textwidth]{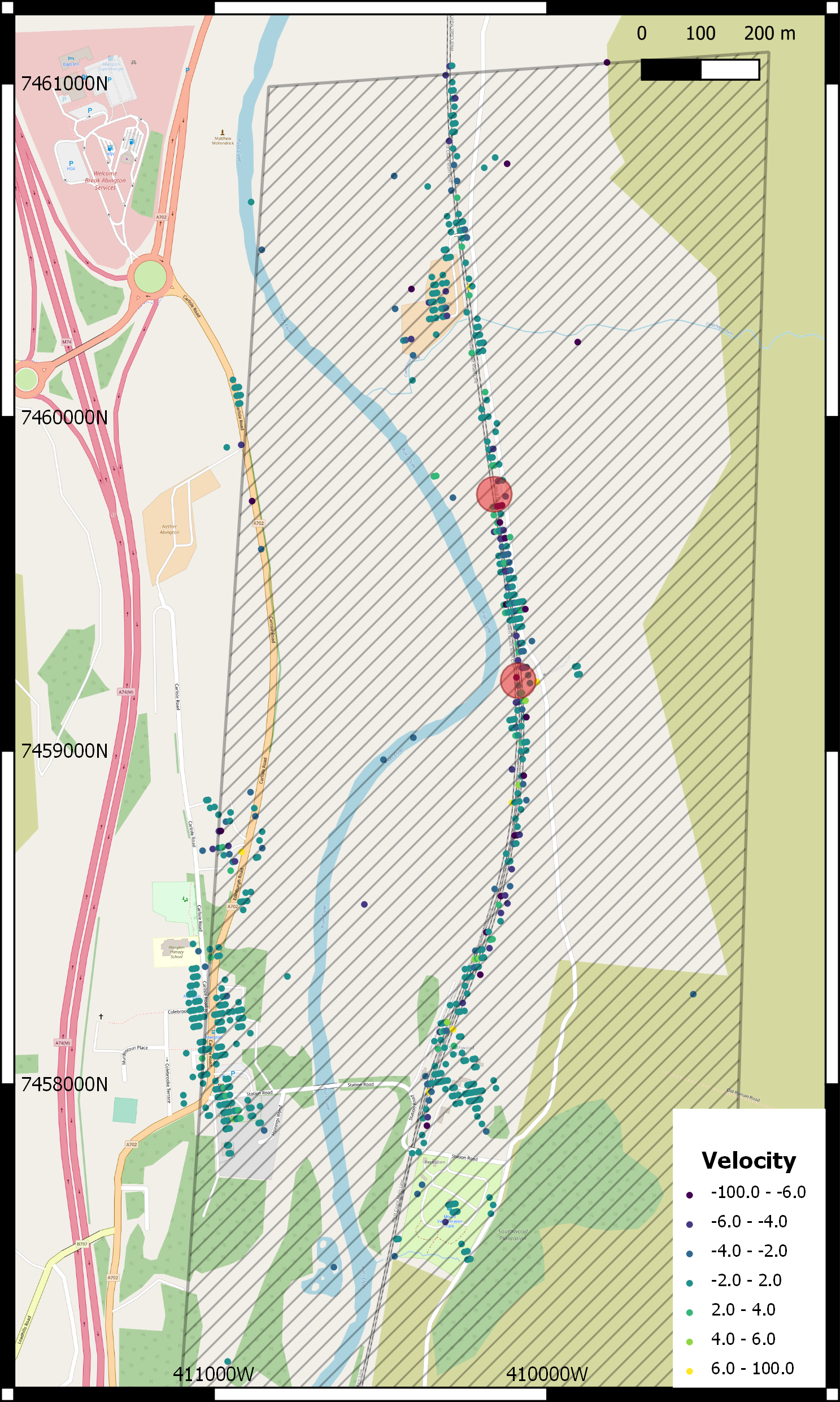}}
\caption{Example of SatSense data from (a) Site A and (b) Site B. Fault sites marked in red, datapoint colour indicates average displacement velocity ($mm/yr$). Note the difference in measurement density between rural and built-up areas.}
\label{fig:density}
\end{figure}

An example of this change in measurement density can be seen in Figure 1 - most returns in rural areas are from the railway infrastructure itself, with little else around it. Unfortunately these are the areas in which rail corridor monitoring is of most interest, as monitoring is easier within urban areas. 

 \subsection{Case studies of Railway faults}
These data are complimented by a database of known railway failures provided by Network Rail. We have identified a couple of areas in which a known fault due to ground movement has occurred within the time period investigated. 
The first site (Site A) investigated is on the East Coast Line, near Oak Tree Farm just north of Stamford, UK. A fault was identified in the area on April 22nd, 2020, involving loss of support due to translational failure on the cutting slope, with similar faults having also occurred in the area on December 18th, 2019 and on June 13, 2018. The second location (Site B) is on the West Coast Main Line, northeast of Abington, UK. Two faults were identified here on January 7th and January 13th, 2016, involving a translational failure of the natural slope by the rail tracks due to excessive rainfall.


\section{PREPROCESSING \& INTERPOLATION}

\subsection{Time Series Interpolation}
Given an area of interest, our dataset will consist of a number of points, unevenly distributed across the scene, each representing a time series of displacement measurements from May 2015 to current day (July 2021 for the data used in this paper). The timeframe selected is of course subject to each scenario of interest. 

The time series data may contain missing values (represented as NaN) at dates when a particular datapoint was too incoherent so as to extract a meaningful signal from it. Water cover, as well as thick vegetation can cause a datapoint to be incoherent at certain dates; a change in the built environment, a landslide or any other major change in the landscape can also cause a previously coherent point to become permanently incoherent past a certain date.

We begin by identifying datapoints that contain too many NaN measurements, typically due to a point becoming permanently incoherent after significant environmental change. These timeseries are deemed unusable and removed from the dataset. For this study we experimentally set the cut-off threshold to $ 15 \% $  - any time series that is not coherent over at least $ 85 \% $ of its total sample length is eliminated.

\begin{figure}[!t]
  \centering
  \centerline{\includegraphics[width=8.5cm]{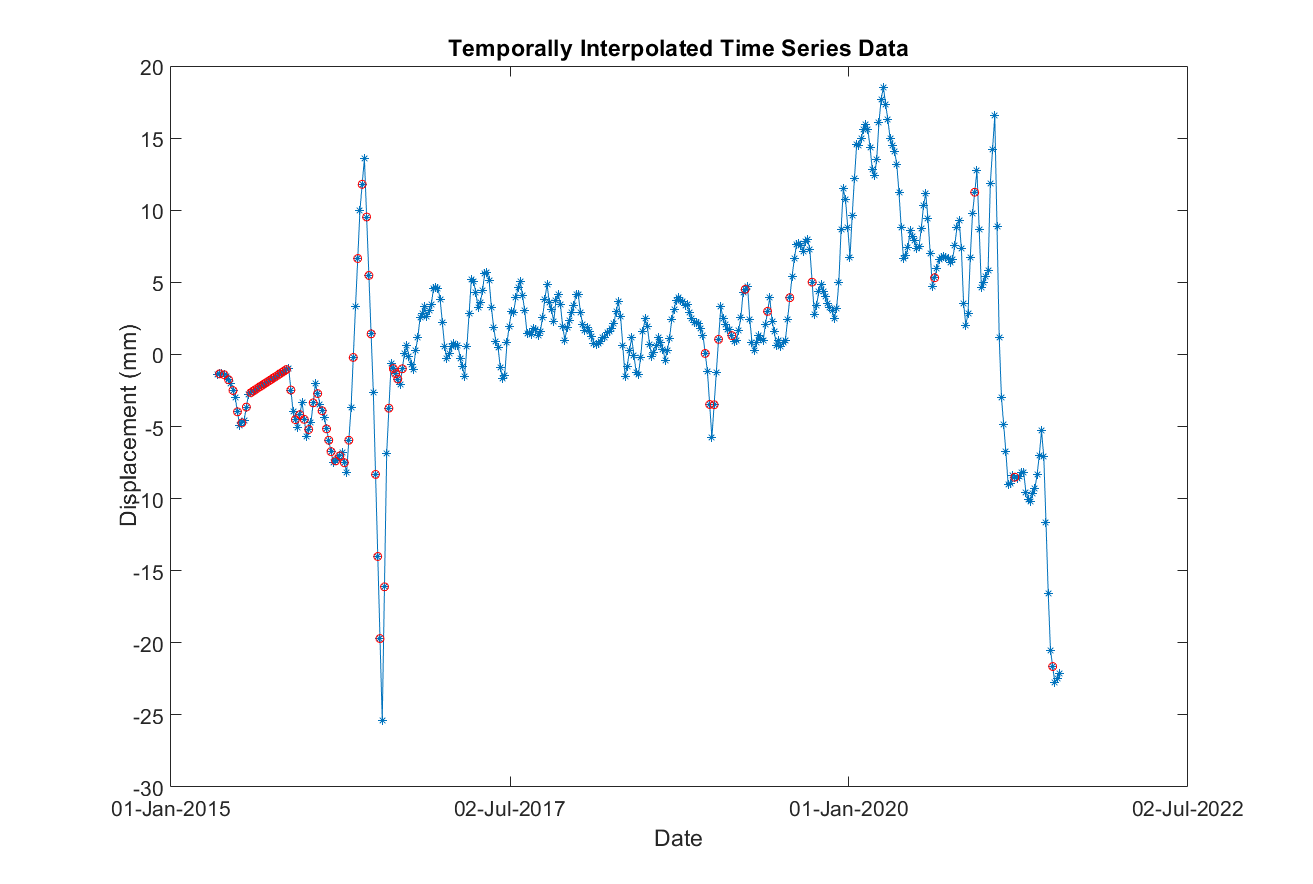}}
\caption{Sample time series data with temporally interpolated points (missing satellite acquisitions) circled in red.}
\label{fig:time_interps}
\end{figure}

The Sentinel-1 constellation has a revisit time of 6 days, but this was only achieved with the launch of Sentinel-1B in April 2016; prior to that the revisit time of Sentinel-1A was at 12 days. Additionally, there are isolated instances of SAR acquisition frames missing from the temporal sequence. All selected time series are therefore amended to a regular sampling period of 6 days, with NaN-valued time frames inserted on these missing dates, e.g. as in Figure \ref{fig:time_interps}.

NaN values in the time series, whether pre-existing or added during the temporal sampling correction, indicate missing data points, and then need to be interpolated. In order to avoid unnecessary assumptions we here opt for a simple linear interpolation between neighbouring frames in the time series.

Finally, we include a simple median denoising step here aimed at mitigating the spiky noise in the time series data. The process applies a 2-D median filter $Med_{3\times3}(\cdot)$ that omits NaN values in the median calculation.  The time series data are then projected into a 3-dimensional sparse image stack.

\subsection{Spatio-temporal Interpolation}

This stack now needs to be interpolated into a dense 3-dimensional volume. We have utilised two inpainting methods to do so. Inpainting typically refers to algorithms for filling in gaps in images, though they can be extended to more general interpolation/extrapolation tasks as in this case. 

Such techniques are often based on solving partial differential equations (PDE) formulated over the image space, typically emulating physical analogues such as plate deformation, spring motion, diffusion or fluid motion \cite{Crema}. These have been found to produce interpolation results that tend to preserve and propagate linear and textural features, a property highly desirable in the context of image inpainting \cite{Chan}. We have here opted for a spring model \cite{DERICO}, where pixels are connected to their neighbours (horizontally and vertically) via springs of infinitesimal length. 

Additionally, we have attempted inpainting using the Deep Image Prior (DIP) framework \cite{DIP}. DIP is a state-of-the-art method capable of addressing a variety of inverse problems in imaging, from denoising to inpainting. It uses a randomly initialised deep neural network as the prior in an inverse problem formulation. Inverse problems are typically formulated as the minimisation of (\ref{eq:inve}) aiming to recover the image $x$ from corrupted observation $x_0$, where $E(x;x_0)$ is the data term and $R(x)$ is a regularisation image prior.

\begin{equation}
min_x E(x;x_0) + R(x)
\label{eq:inve}
\end{equation}

The DIP framework instead attempts to solve this problem by minimising 
\begin{equation}
min_{\theta} E(f_{\theta}(z);x_0)
\label{eq:DIP}
\end{equation}

\noindent where $f$ is a deep convolution neural network with parameters $\theta$ and $z$ is a fixed noise input. This means the minimisation solution is searched not on the image space but on the space of the deep network's parameters.

The network used here is U-net architecture with a depth of 3. All depths have 128 feature channels, upsampling is by bilinear interpolation and the activation layer is a leaky ReLU. We use the Adam optimiser, with an initial learning rate of 0.001. As the dataset here is quite small, the process terminates after 3000 epochs, otherwise obvious overfitting is observed.




\section{EXPERIMENTAL RESULTS }

\begin{figure}[!t]
\centering
\subfloat[]{\includegraphics[width=0.15\textwidth]{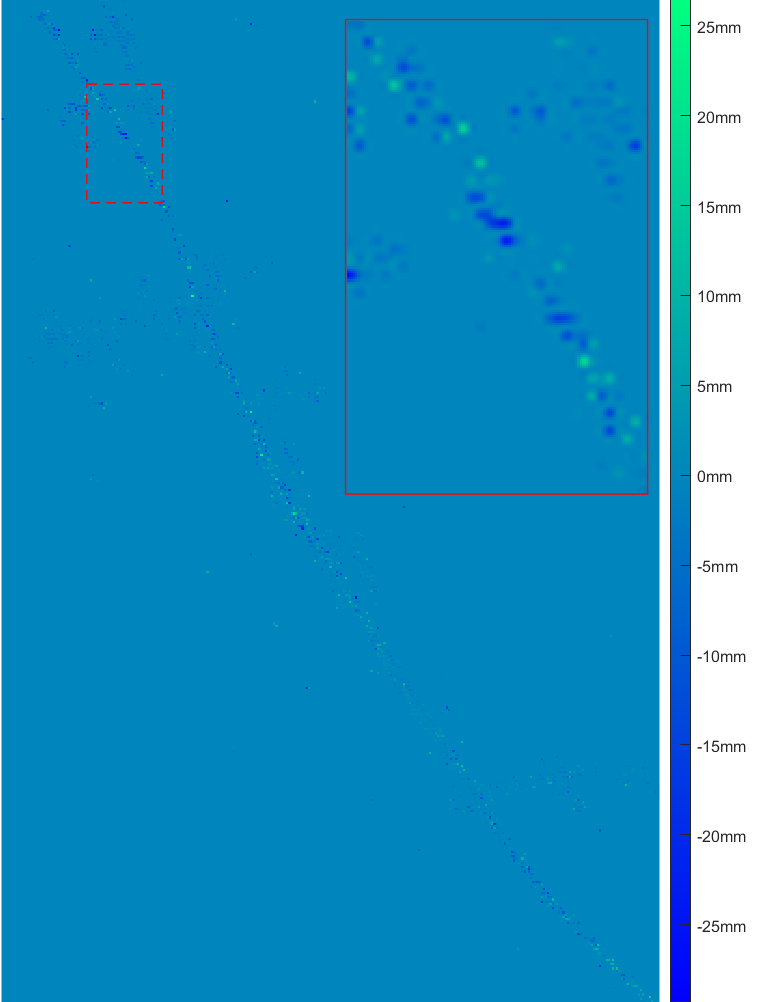}}
\hspace{0.15cm}
\subfloat[]{\includegraphics[width=0.15\textwidth]{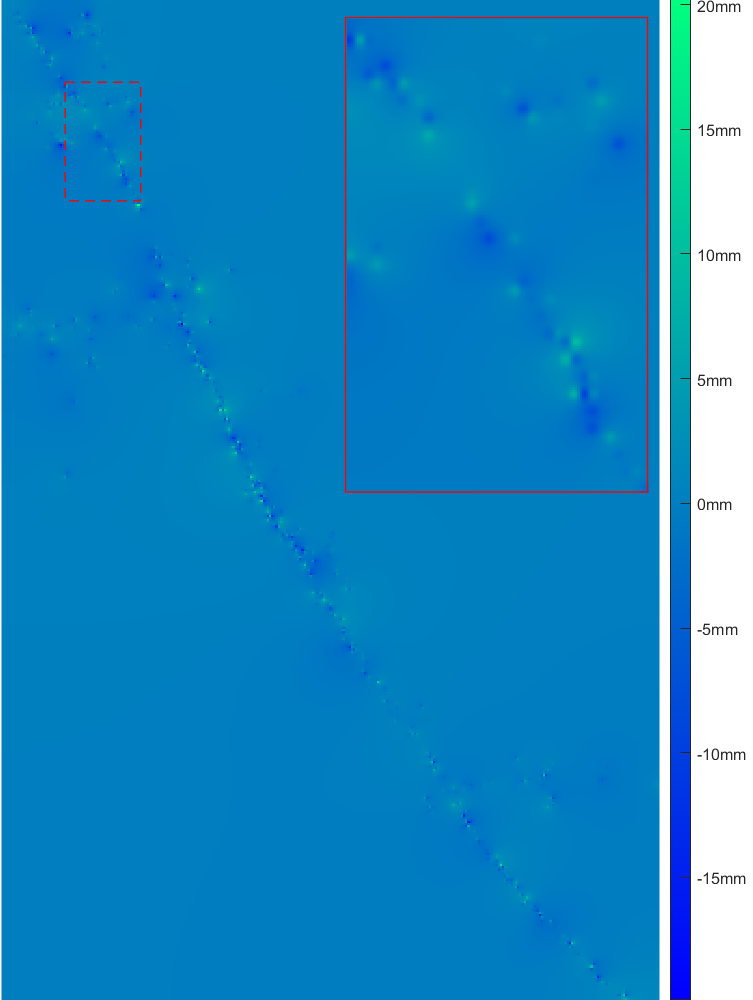}}
\hspace{0.15cm}
\subfloat[]{\includegraphics[width=0.15\textwidth]{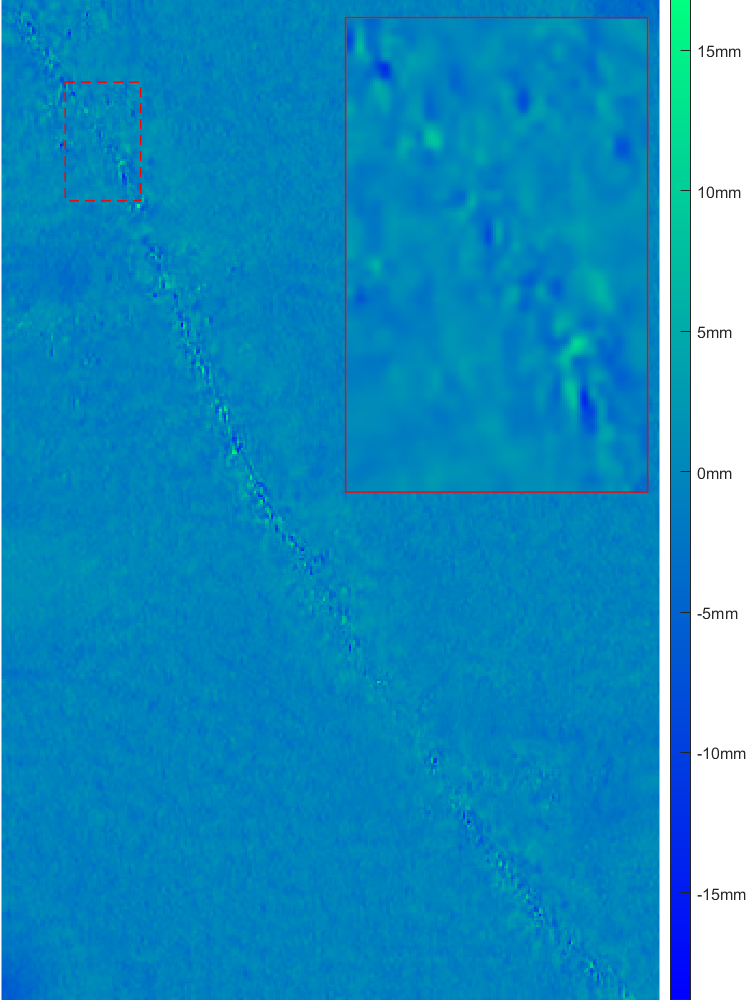}}
\caption{Site A (a) Raw data (random stack sample), corresponding (b) PDE 3D interpolated, (c) DIP 3D interpolated data. Top Right: magnified area of known faults.}
\label{fig:114}
\end{figure}

\begin{figure}[!t]
\centering
\subfloat[]{\includegraphics[width=0.15\textwidth]{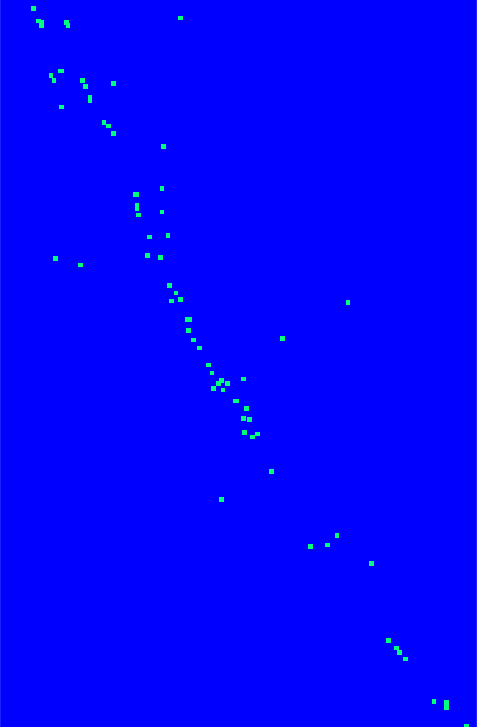}}
\hspace{0.05cm}
\subfloat[]{\includegraphics[width=0.15\textwidth]{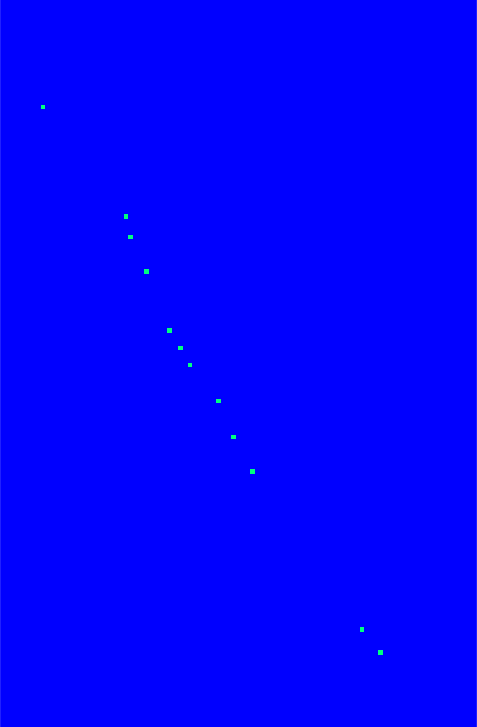}}
\hspace{0.05cm}
\subfloat[]{\includegraphics[width=0.15\textwidth]{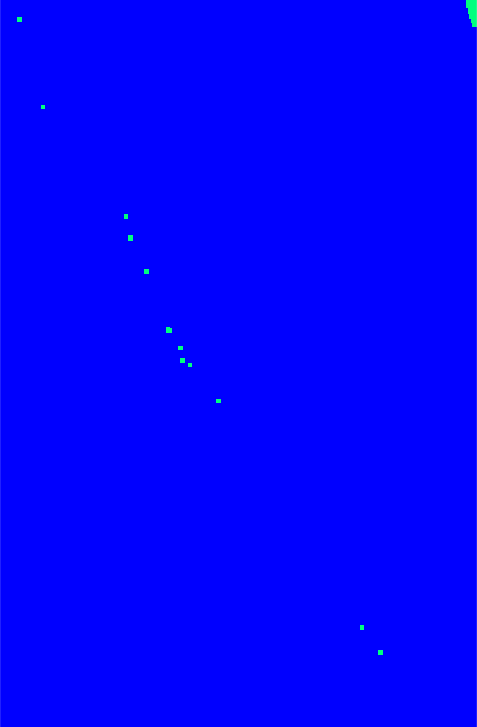}}
\caption{Binary detection maps of points in Site A exhibiting high variance ($>150mm^2$) derived from (a) untreated, sparse displacement data, (b) proposed method using PDE inpainting and (c) proposed method using DIP inpainting. Detections dilated for display purposes.}
\label{fig:114_var_simple}
\end{figure}

A simple demonstration of the benefits of working with the interpolated, denser data as opposed to the original raw data can be seen in Figures \ref{fig:114} \& \ref{fig:114_var_simple}. The data as originally provided contains irregularities and noise that effectively necessitates treatment as described in Section 3.1 prior to any further processing. The data is however still very noisy at that point, with individual points demonstrating high variance over time. Interpolating the time series into dense image stacks can reduce this apparent variance in the data, making significant movements easier to discern from the noise.  Figure \ref{fig:114_var_simple} shows binary detection maps of points in the scene that demonstrate variance above an experimentally set threshold of $150mm^2$. Variance here is calculated per point over the entire time series length / image stack depth. 

While the spatio-temporally interpolated stacks contain far more points of measurable variance compared to the sparse raw data stack, a considerably smaller number of those exhibits high variance; additionally, these points (as will be seen later) correlate with known faults in the area.

\begin{figure}[!t]
\centering
\subfloat[13/06/2018]{\includegraphics[width=0.15\textwidth]{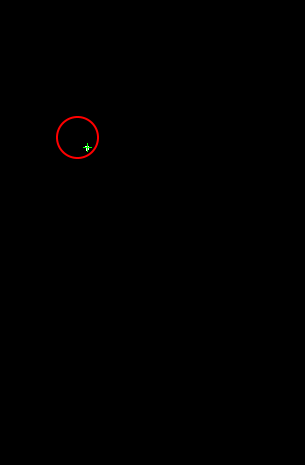}}
\hspace{0.15cm}
\subfloat[23/12/2019]{\includegraphics[width=0.15\textwidth]{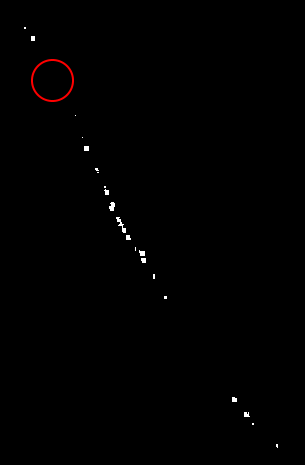}}
\hspace{0.15cm}
\subfloat[27/04/2020]{\includegraphics[width=0.15\textwidth]{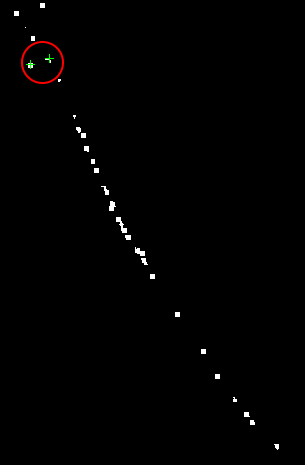}}

\subfloat[13/06/2018]{\includegraphics[width=0.15\textwidth]{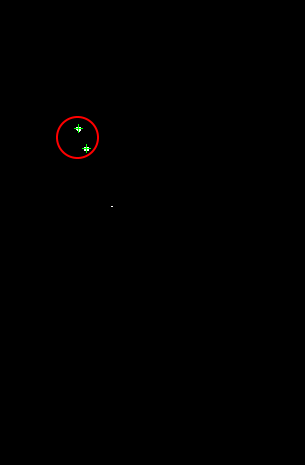}}
\hspace{0.15cm}
\subfloat[23/12/2019]{\includegraphics[width=0.15\textwidth]{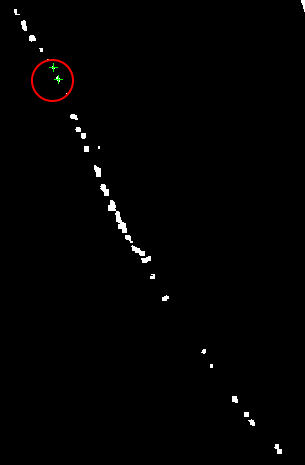}}
\hspace{0.15cm}
\subfloat[27/04/2020]{\includegraphics[width=0.15\textwidth]{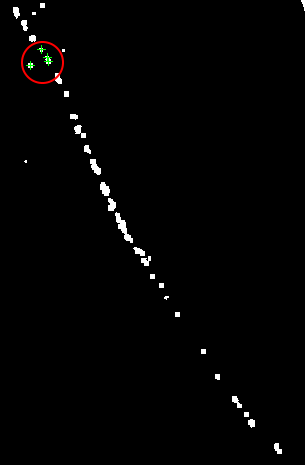}}
\caption{Local Variance Detector results for Site A, first available frame after fault dates. (a), (b) and (c) detector output on PDE interpolation, (d), (e), (f) detector output on DIP interpolation. Fault area (aprox.) marked in red, positive detections marked green. Variance threshold at $50mm^2$, calculated over $5\times5\times5$ window. }
\label{fig:114exp}
\end{figure}

\begin{figure}[h]
\centering
\subfloat[]{\includegraphics[width=0.15\textwidth]{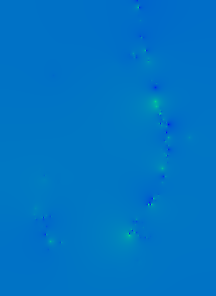}}
\hspace{0.15cm}
\subfloat[]{\includegraphics[width=0.15\textwidth]{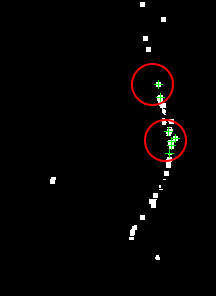}}

\subfloat[]{\includegraphics[width=0.15\textwidth]{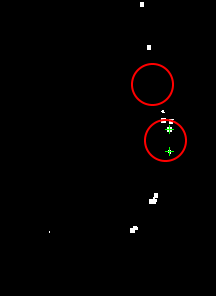}}
\hspace{0.15cm}
\subfloat[]{\includegraphics[width=0.15\textwidth]{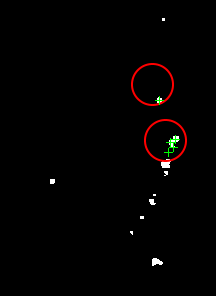}}
\caption{Local Variance Detector results for Site B. (a) raw data, (b) detector output on PDE interpolation, (c) detector output on DIP interpolation. Fault locations (approx.) marked in red,positive detections marked green. Variance threshold at $5mm^2$, calculated over $5\times5\times5$ window. Frame taken at 18th Jan. 2016.}
\label{fig:107exp}
\end{figure}

For a more practical experiment, we redesigned our detector to compute local variance in a $5x5$ neighbourhood over approximate month intervals, i.e. 5 frames in the temporal dimension. Figure \ref{fig:114exp} shows the binary thresholding result of the variance map on 3 dates corresponding to the 3 known faults in Site A. The frames shown are the first available after the date on which the fault was reported, which may of course not necessarily coincide with an InSAR acquisition date for the site. The fault area has been marked in a red circle, and it can be seen that this simple detector is capable of capturing motion in that region in most cases.

The DIP variant performs slightly better for the same variance threshold, and detects movement that can reasonably be attributed with a fault (Figure \ref{fig:114exp} (e)) that the PDE variant misses. Note there may well be motion detected by our method in parts of the image that does not correspond with a known fault, simply because no fault resulted of it (especially when this movement did not occur at a point near the rail corridor). Instead we are interested in seeing whether our method can detect in the data ground movement that can reasonably be correlated, spatially and temporally, to the failures known.

We repeat a similar experiment for Site B, shown in Figure \ref{fig:107exp}. As both known faults here occurred within a few days, they are marked on the same frame. We also include an example of the detector on the raw data, demonstrating the much higher number of detections, almost equal to the number of total datapoints.

\section{Conclusions and Future Work}

We demonstrate here a scheme for treating noisy, sparse InSAR data and interpolating them into a dense spatiotemporal image stack, along with a proof of concept experiment for the detection of significant ground movement near the rail corridor. We wish to extend this work in the near future to further improve data conditioning, develop a novel motion detector using more advanced deep learning methods and to conduct a larger scale field study for more accurate correlation of ground movement and railway faults.

\vfill\pagebreak


\clearpage
\balance

\bibliographystyle{IEEEtran}
\bibliography{ICIPrailway}

\begin{thebibliography}{10}
\providecommand{\url}[1]{#1}
\csname url@samestyle\endcsname
\providecommand{\newblock}{\relax}
\providecommand{\bibinfo}[2]{#2}
\providecommand{\BIBentrySTDinterwordspacing}{\spaceskip=0pt\relax}
\providecommand{\BIBentryALTinterwordstretchfactor}{4}
\providecommand{\BIBentryALTinterwordspacing}{\spaceskip=\fontdimen2\font plus
\BIBentryALTinterwordstretchfactor\fontdimen3\font minus
  \fontdimen4\font\relax}
\providecommand{\BIBforeignlanguage}[2]{{%
\expandafter\ifx\csname l@#1\endcsname\relax
\typeout{** WARNING: IEEEtran.bst: No hyphenation pattern has been}%
\typeout{** loaded for the language `#1'. Using the pattern for}%
\typeout{** the default language instead.}%
\else
\language=\csname l@#1\endcsname
\fi
#2}}
\providecommand{\BIBdecl}{\relax}
\BIBdecl

\bibitem{WOODBURN}
A.~Woodburn, ``{Rail Network Resilience and Operational Responsiveness During
  Unplanned Disruption: A Rail Freight Case Study},'' \emph{Journal of
  Transport Geography}, vol.~77, pp. 59--69, 2019.

\bibitem{CHANG}
L.~Chang, R.~P. B.~J. Dollevoet, and R.~F. Hanssen, ``{Nationwide Railway
  Monitoring Using Satellite SAR Interferometry},'' \emph{IEEE Journal of
  Selected Topics in Applied Earth Observations and Remote Sensing}, vol.~10,
  no.~2, pp. 596--604, 2017.

\bibitem{Tofani}
V.~Tofani, F.~Raspini, F.~Catani, and N.~Casagli, ``{Persistent Scatterer
  Interferometry (PSI) Technique for Landslide Characterization and
  Monitoring},'' \emph{Remote Sensing}, vol.~5, no.~3, pp. 1045--1065, 2013.

\bibitem{NORTH}
M.~North, T.~Farewell, S.~Hallett, and A.~Bertelle, ``{Monitoring the Response
  of Roads and Railways to Seasonal Soil Movement with Persistent Scatterers
  Interferometry over Six UK Sites},'' \emph{Remote Sensing}, vol.~9, no.~9,
  2017.

\bibitem{PUI1}
N.~Anantrasirichai, J.~Biggs, K.~Kelevitz, Z.~Sadeghi, T.~Wright, J.~Thompson,
  A.~M. Achim, and D.~Bull, ``{Detecting Ground Deformation in the Built
  Environment Using Sparse Satellite InSAR Data With a Convolutional Neural
  Network},'' \emph{IEEE Transactions on Geoscience and Remote Sensing},
  vol.~59, no.~4, pp. 2940--2950, 2021.

\bibitem{HILL}
P.~Hill, J.~Biggs, V.~Ponce-López, and D.~Bull, ``{Time-Series Prediction
  Approaches to Forecasting Deformation in Sentinel-1 InSAR Data},''
  \emph{Journal of Geophysical Research: Solid Earth}, vol. 126, no.~3, p.
  e2020JB020176, 2021.

\bibitem{PUI2}
N.~Anantrasirichai, J.~Biggs, F.~Albino, and D.~Bull, ``{The Application of
  Convolutional Neural Networks to Detect Slow, Sustained Deformation in InSAR
  Time Series},'' \emph{Geophysical Research Letters}, vol.~46, no.~21, pp.
  11\,850--11\,858, 2019.

\bibitem{InSAR2018}
N.~Anantrasirichai, J.~Biggs, F.~Albino, P.~Hill, and D.~Bull, ``{Application
  of Machine Learning to Classification of Volcanic Deformation in
  Routinely-Generated In{SAR} data},'' \emph{J. Geophys. Res.: Solid Earth},
  vol. 123, no.~8, pp. 6592--6606, August 2018.

\bibitem{RapidSAR}
K.~Spaans and A.~Hooper, ``{InSAR Processing for Volcano Monitoring and Other
  Near-real Time Applications},'' \emph{Journal of Geophysical Research: Solid
  Earth}, vol. 121, no.~4, pp. 2947--2960, 2016.

\bibitem{Crema}
S.~Crema, M.~Llena, A.~Calsamiglia, J.~Estrany, L.~Marchi, D.~Vericat, and
  M.~Cavalli, ``{Can Inpainting Improve Digital Terrain Analysis? Comparing
  Techniques for Void Filling, Surface Reconstruction and Geomorphometric
  Analyses},'' \emph{Earth Surface Processes and Landforms}, vol.~45, no.~3,
  pp. 736--755, 2020.

\bibitem{Chan}
T.~F. Chan and J.~Shen, ``{Nontexture Inpainting by Curvature-Driven
  Diffusions},'' \emph{Journal of Visual Communication and Image
  Representation}, vol.~12, no.~4, pp. 436--449, 2001.

\bibitem{DERICO}
{J. D'Erico}, ``{Inpaint NaNs},'' MATLAB File Exchange, 2006, accessed 20th
  Nov., 2021.

\bibitem{DIP}
D.~Ulyanov, A.~Vedaldi, and V.~Lempitsky, ``{Deep Image Prior},''
  \emph{International Journal of Computer Vision}, vol. 128, pp. 1867--1888,
  2020.

\end{thebibliography}

\end{document}